\documentclass[12pt]{article} 
\usepackage{graphicx}
\usepackage{amsmath}
\usepackage{amssymb}
\usepackage{txfonts} 
\usepackage[left=2cm,right=2cm]{geometry}
\usepackage{setspace} 
\newcommand\fverb{\setbox\pippobox=\hbox\bgroup\verb}
\newcommand\fverbdo{\egroup\medskip\noindent%
			\fbox{\unhbox\pippobox}\ }
\newcommand\fverbit{\egroup\item[\fbox{\unhbox\pippobox}]}
\newbox\pippobox

\def\be{\begin{equation}}
\def\ee{\end{equation}}
\def\ba{\begin{array}{l}}
\def\ea{\end{array}}
\def\bea{\begin{eqnarray}}
\def\eea{\end{eqnarray}}
\def\beas{\begin{eqnarray*}}
\def\eeas{\end{eqnarray*}}
\def\eq#1{(\ref{#1})}
\def\fig#1{Figure \ref{#1}}

\begin{document}
\title{Intersecting branes and Nambu$-$Jona-Lasinio model}
\author{Avinash Dhar and Partha Nag \\  
Tata Institute of Fundamental Research, Homi Bhabha Road \\ 
Mumbai 400 005, India 
}


\maketitle

\enlargethispage*{1000pt}    
\abstract
\fontsize{12pt}{15pt}\selectfont
{We discuss chiral symmetry breaking in the intersecting brane model
of Sakai and Sugimoto at weak coupling for a generic value of
separation $L$ between the flavour $D8$ and anti-$D8$-branes. For any
finite value of the radius $R$ of the circle around which the colour
$D4$-branes wrap, a non-local Nambu$-$Jona-Lasinio (NJL) type
short-range interaction couples the flavour branes and anti-branes. We
argue that chiral symmetry is broken in this model only above a
certain critical value of the $4$-dimensional 't Hooft coupling and
confirm this through numerical calculations of solutions to the gap
equation. We also numerically investigate chiral symmetry breaking in
the limit $R \rightarrow \infty$ keeping $L$ fixed, but find that
simple ways of implementing this limit do not lead to a consistent
picture of chiral symmetry breaking in the non-compact version of the
non-local NJL model.}



\newpage

\tableofcontents    
\section{Introduction}

The Nambu$-$Jona-Lasinio (NJL) model \cite{NJL} provides an example of
dynamical chiral symmetry breaking ($\chi$SB) and fermion mass
generation in a simple effective field theory setting. In the original
model, the fermions in the four-fermi interaction were taken to be the
nucleons. Interest in the model has endured for two main reasons: (i)
It appears to give a rather accurate description of chiral symmetry
breaking and its consequences for low-energy hadron phenomenology;
(ii) Appropriately replacing the original nucleons by coloured quarks,
the model can be argued to describe all of the low-energy physics of
QCD, including the anomaly term \cite{DW,DSW,D}
\footnote{These works give an argument based on Wilsonian RG and the
confinement property of QCD for the emergence of NJL model for quarks
from the underlying microscopic dynamics, including the correct
anomaly term with a coefficient proportional to the number of
colours. For a review of applications of this model to QCD
phenomenology, see \cite{HK}}.

Recently, versions of the NJL model have emerged in a string theory
setting \cite{AHJK,AHK1}, involving intersecting brane
configurations. One such configuration is the model of Sakai and
Sugimoto (SS) \cite{SS1}, which involves a system of intersecting
$D4$, $D8$ and anti-$D8$-branes. The SS model has been very successful
in reproducing many of the qualitative features of non-abelian chiral
symmetry breaking in QCD. In this model, the `colour' Yang-Mills
fields are provided by the massless open string fluctuations of a
stack of $N_c$ $D4$-branes, which are extended along the four
space-time directions and in addition wrap a thermal circle of radius
$R$. At scales much larger than string length, the theory on the
$D4$-branes is $(4+1)$-dimensional pure Yang-Mills with coupling
$g_5^2=(2\pi)^2 g_s l_s$ of length dimension. In the strong coupling
limit, $g_5^2 N_c  >> 2\pi R$, this stack of
$D4$-branes has a dual description in terms of a classical gravity
theory \cite{EW} with the background geometry of a Euclidean black
hole. Flavour degrees of freedom \cite{KK,BEEGK,KMMW} are provided by
the massless open string fluctuations between the colour branes and
the `flavour' $D8$ and anti-$D8$-branes, which intersect the thermal
circle at points separated by a distance $L \leq \pi R$. Various
aspects of chiral symmetry breaking in this model have been discussed
in \cite{SS1,SS2,AHJK,AHK1,NSK,NSK2,HRYY,HSSY,BLL,DY,ASY,PS0,CKP,BSS,DN1,
DN2,AK,HHLY,HSS,ELN,MMS,AELV}.

It was pointed out in \cite{AHJK} that the brane configuration of the
SS model decouples the scales of chiral symmetry breaking and
confinement \footnote{A similar observation was made in \cite{BY} in a
different context.}. The additional parameter in the SS model (as
compared to QCD) which makes this possible is the ratio $L/R$. The
authors of \cite{AHJK} argued that in the limit $R \rightarrow \infty$
with $L$ kept fixed (non-compact SS model), the effective low-energy
description of the SS model at weak coupling \footnote{There are
several parameters of length dimension in the SS model, viz. $R$, $L$,
$l_s$ and $g_5^2$. As discussed in \cite{AHJK}, the weak coupling
limit is defined by the hierarchy of scales $g_5^2 N_c << l_s << L <<
R$. In this parameter region, stringy effects may be neglected and, as
we shall see, a controlled treatment of the interaction between left
and right-handed flavours, mediated by Yang-Mills fields, can be
given. The condition $L << R$ makes it possible to have a chiral
symmetry breaking (length) scale which is much smaller than the
confinement scale.} is given by a non-local version of the NJL model,
which breaks chiral symmetry spontaneously at arbitrarily weak
coupling. This result is surprising in view of our field theoretic
intuition, which would suggest that chiral symmetry is not broken at
weak coupling and that there is a transition to the broken phase at
some critical value. One might suspect that this unexpected result is
connected with the absence of a mass gap in the non-compact SS model,
which results in a long-range four-fermi interaction. One way to test
this hypothesis would be to work with a finite, but possibly very
large, value of $R$, which corresponds to a confining theory with
possibly a small, but non-zero, mass gap. From general arguments, one
would expect chiral symmetry to be broken in this model at a length
scale of the order of or smaller than the confinement scale
\footnote{It is generally believed that in a confining gauge theory 
the length scale associated with $\chi$SB is of the order of or
smaller than the confinement length scale. QCD is an example where the
$\chi$SB scale is of the order of the confinement scale.
Recently, it has been argued in \cite{ASY,PS0} that in the strongly
coupled SS model the $\chi$SB length scale can be smaller than the
confinement length scale, depending on the value of $L$.}.  However,
in the corresponding effective non-local NJL model which describes the
flavour brane-antibrane interactions, $\chi$SB could be associated
with length scales even larger than the confinement scale. This is
because the NJL model does not incorporate confinement. Now, the point
is that for solutions which have $\chi$SB scale larger than the
confinement scale, an effective local NJL model should be
adequate. But this latter model shows a critical coupling for
$\chi$SB. This argument suggests that an effective non-local NJL model
for the SS model would show $\chi$SB only beyond a critical value of
the coupling.

In this paper we analyze $\chi$SB at weak coupling in the brane
configuration of the SS model, with a finite value of $R$ and a
generic value of $L$. This theory has a mass gap, which gives the
scale over which the four-fermi coupling of the flavour branes
extends. We obtain the leading order (in gauge coupling) approximation
to the effective fermion action, including the exact contribution due
to the Kaluza-Klein modes. We expand on the plausibility argument
given above that in the resulting non-local NJL model chiral symmetry
is spontaneously broken only above a certain critical coupling. We
then verify this by obtaining numerical solutions to the gap equation
derived from the effective fermi theory. The plan of this paper is as
follows. In the next section we first briefly review the argument of
\cite{DSW} for the emergence of local NJL model from QCD and then
extend it to a non-local model. We use this in Section 3 to derive the
non-local NJL model as the leading approximation to the coupling of
the flavour branes in the weakly coupled SS model. In Section 4, we
discuss $\chi$SB in the non-local NJL model. We derive the gap
equation in the large $N_c$ limit and present numerical solutions
which show that chiral symmetry is spontaneously broken only above a
certain critical coupling. A discussion of the non-compact case is
given in Section 5. We end with a summary in Section 6.

\section{\label{njlqcd}NJL model from QCD}

The Yang-Mills action for $U(N_c)$ gauge group (indices $a,b$) and
$N_f$ massless quark flavours (indices $\alpha,\beta$) is
\footnote{We use the following notations and conventions. The space-time 
metric is mostly minus. Our Dirac matrices and their Weyl
representation are as given in \cite{PS}, in particular, the equations
(3.41) and (3.42). We have used the notation $x^\mu$ ($\mu=0,1,2,3$)
to label the four space-time coordinates. Also, $t^a$ are hermitian
generators of $U(N_c)$ in the fundamental representation. In
particular, we will need the identity
$(t^a)_{ij}(t^a)_{kl}=\frac{1}{2}\delta_{il}\delta_{jk}$.}
\bea
S_0=-\frac{1}{4g_4^2}\int d^4x \ (F_{\mu\nu}^a(x))^2+ \int d^4x
\ {\bar q}^\alpha(x) \gamma^\mu \biggl(i\partial_\mu+
t^a A^a_\mu(x)\biggr)q^\alpha(x).
\label{ea2}
\eea
This theory confines and develops a mass scale $\Lambda$, given by
\footnote{This is true for $N_f < 11N_c/2$, which is easily satisfied
in the large $N_c$ and fixed $N_f$ limit that we will be interested in
here. Also, the mass scale $m$ that enters in this formula should be
taken to be the scale at which the input coupling $g_4^2$ is
measured.}
\bea
\Lambda \sim m \ e^{-\frac{1}{\beta_0g_4^2}}, 
\qquad \beta_0=\frac{1}{24\pi^2}(11N_c-2N_f).
\label{scale}
\eea
It is generally believed that at energies below the confining scale,
an effective NJL model for quarks captures the dynamics of the
theory. There is no systematic way of integrating out the Yang-Mills
degrees of freedom from QCD to get an effective fermion action. A
scenario outlining how one might think about doing this was presented
in \cite{DSW}. The basic point is that integration of Yang-Mills
degrees of freedom would lead to effective multi-quark
interactions. The range of these interactions must be short, of the
order of $1/\Lambda$, because of confinement, and so at energies below
$\Lambda$ a local approximation would be adequate. The NJL interaction
between gauge-invariant quark bilinears is the leading term compatible
with gauge symmetry and global symmetries of QCD.

\subsection{\label{nlnjl}Extension to a non-local NJL model}

In QCD it is generally believed that the mass scale $M$ associated
with $\chi$SB coincides with the confinement scale $\Lambda$. Suppose,
however, we can deform QCD in such a way that the two scales are
separated by some new physics (as in the SS intersecting brane
configuration discussed in the next section). In this case, for
studying $\chi$SB we need the effective four-fermi theory at energies
larger than the mass gap $\Lambda$. If $\Lambda << M$, the energies of
quarks involved in the four-fermi interaction are much larger than
$\Lambda$. Because of asymptotic freedom, for energies much larger
than $\Lambda$ we can present a more precise derivation of the
effective interaction. The leading contribution to the effective
interaction comes from a one-gluon exchange approximation, which can
be calculated exactly. The result
\footnote{As usual, to do the calculation one needs to fix a gauge. 
The calculation done here and in the next section uses the Feynman
gauge.} is
\bea
S_{0~{\rm eff}}=-\frac{g_4^2}{2} \int d^4x \ d^4y \ \Delta_0(x-y) \
J^{a\mu}(x)J^a_\mu(y)
\label{ea22}
\eea
where
\bea
J^{a\mu}(x)=\biggl(q^\dagger_L(x) {\bar \sigma}^\mu t^a q_L(x)+
q^\dagger_R(x) \sigma^\mu t^a q_R(x)\biggr),
\label{j0} 
\eea
and
\bea
\Delta_0(x)=\int \frac{d^4k}{(2\pi)^4} \frac{e^{ik.x}}{{\bar k}^2},
\qquad  {\bar k}^2 \equiv {\vec k}^2-k_0^2-i\epsilon.
\label{g01}
\eea
is the Feynman propagator for a massless scalar \footnote{This rather
unfamiliar way of writing the Feynman propagator is convenient since
on making a Wick rotation to Euclidean signature and setting
$\epsilon$ to zero, $\bar k$ becomes just the magnitude of the
Euclidean $4$-momentum $k_\mu$.}. Using the Fierz identities given in
equations (3.77) and (3.80) of \cite{PS} and retaining only
interaction terms between left and right-handed Weyl components
\footnote{The four-fermi terms involving Weyl components of a single 
handedness are not relevant to the discussion of chiral symmetry
breaking vacuum. Hence, these terms are not taken into account here.},
the effective action \eq{ea22} becomes
\bea
S_{0~{\rm eff}}=g_4^2 \int d^4x \ d^4y \ \Delta_0(x-y)
[q^{\dagger\alpha}_L(x)q^\beta_R(y)][q^{\dagger\beta}_R(y)q^\alpha_L(x)],
\label{ea23}
\eea
The bilocal fermion products in square brackets are singlets of the
(global) colour $U(N_c)$ group and transform as $({\bar N_f},N_f)$ under
the flavour $U(N_f) \times U(N_f)$ group.

The above discussion may be summarized as follows. The effective
four-fermi theory resulting from integrating out the gluon degrees of
freedom is
\bea
S_{0~{\rm eff}} &=& i\int \ d^4x \biggl(q^{\dagger\alpha}_L(x) 
{\bar \sigma}^\mu \partial_\mu q^\alpha_L(x)+
q^{\dagger\alpha}_R(x) \sigma^\mu \partial_\mu q^\alpha_R(x)\biggr)
\nonumber \\
&& +g_4^2 \int d^4x \ d^4y \ G_0(x-y)
[q^{\dagger\alpha}_L(x)q^\beta_R(y)][q^{\dagger\beta}_R(y)q^\alpha_L(x)],
\label{ea24}
\eea
where
\bea
G_0(x)=\int \frac{d^4k}{(2\pi)^4} \ e^{ik.x} \ \tilde G_0(k),
\label{g02}
\eea
$\tilde G_0(k)$ satisfies:

\begin{itemize} 

\item For ${\bar k} << \Lambda$, $\tilde G_0(k) \sim$ constant. This is 
ensured by the property of confinement of the action \eq{ea2}, which
leads to the generation of a mass gap. The constant has length
dimension two and we may take it to be $1/\Lambda^2$. In a sense,
this provides a definition of the mass scale $\Lambda$ for us.

\item For $\Lambda << {\bar k} << M$, $\tilde G_0(k) \sim 1/{\bar k}^2$. 
This follows from the asymptotic freedom property of the action
\eq{ea2} and the added new physics which separates the $\chi$SB scale 
from the mass gap of \eq{ea2}.

\end{itemize}

\noindent A simple example of a function $\tilde G_0(k)$ which satisfies 
these two properties is
\bea
\tilde G_0(k)=\frac{1}{{\bar k}^2+\Lambda^2}.
\label{g03}
\eea
A cut-off scale of order $M$ on \eq{g03} is understood.  A more
complicated function could be devised to take into account the running
of the coupling. In any case, the process of integrating out gluon
degrees of freedom in a confining theory is expected to give rise to a
far more complicated effective fermion action than in the model given
by \eq{ea24}, \eq{g02} and \eq{g03}. However, one might hope that the
essential features for studying qualitative questions about $\chi$SB
are present in this model. In the next section, we will see that a
very similar non-local NJL interaction between the flavour branes
emerges as the leading approximation to the weakly coupled SS model.

\section{\label{njlss}NJL model from weakly coupled SS model}

At scales much smaller than the string length, the dynamics of the
weakly coupled SS model is governed by the action \footnote{There are
several possible corrections to this low energy action. For $g_5^2N <<
l_s$ corrections from string modes are small and may be
neglected. Corrections from the string winding modes around the
thermal circle may be neglected for $R >> l_s$. We will assume this to
be the case in the rest of this paper. The low energy effective action
also has possible terms that couple the fermions to the transverse
scalars. However, since the scalars come with a derivative, their
effect may be neglected at low energies.}
\bea
S &=& -\frac{1}{4g_5^2}\int d^4x \int_0^{2\pi R} \ dx^4 \ 
(F_{MN}^a(x,x^4))^2  
+\int \ d^4x \ q^{\alpha\dagger}_L(x) {\bar \sigma}^\mu 
\biggl(i\partial_\mu+t^a A^a_\mu(x,-L/2)\biggr)q^\alpha_L(x) \nonumber \\
&& \hspace{4cm} +\int \ d^4x \ q^{\alpha\dagger}_R(x) \sigma^\mu 
\biggl(i\partial_\mu+t^a A^a_\mu(x,L/2)\biggr)q^\alpha_R(x).
\label{action}
\eea
Only the space-time components $A_\mu(x,\mp L/2)$ \footnote{In
addition to the notations and conventions listed in Footnote $5$, we
use the following conventions. We use $x^4$ to label the coordinate
along the circle which the $D4$-branes wrap. We choose the mid-point
between the locations of the $D8$ and anti-$D8$-branes on the circle,
which are a distance $L$ apart, as the origin in $x^4$. The values
$x^4=\mp L/2$ are then the locations of the $D8$ and anti-$D8$-branes
on the circle.} of the $(4+1)$-dimensional $U(N_c)$ gauge field
$A_M(x,x^4)$ interact with the massless Weyl fermions $q^\alpha_{L,R}(x)$.
Substituting the Kaluza-Klein expansion
\bea
A^a_M(x,x^4)=A_M^{a(0)}(x)+\sum_{n=1}^\infty \biggl(A_M^{a(n)}(x) 
e^{in x^4/R}+{A_M^{a(n)}}^*(x) e^{-in x^4/R}\biggr)\label{gf}
\eea
in this action, we get
\bea
S=S_0+S_1+\cdots
\label{ea1}
\eea
$S_0$ becomes identical to the action \eq{ea2}, after identifying the
gauge potential $A^a_\mu(x)$ of the latter with the zero mode
$A^{a(0)}_\mu(x)$ and setting 
\bea
g_4^2=g_5^2/2\pi R. 
\label{YMg}
\eea
Also, it is now natural to identify the mass scale $m$ in \eq{scale}
with $1/\pi R$ since the $4$-dimensional description breaks down
beyond this scale. $S_1$ is given by
\bea
S_1 &=& \frac{1}{g_4^2} \sum_{n=1}^\infty \int d^4x 
\biggl(-\frac{1}{2} |\partial_\mu A_\nu^{a(n)}(x)
-\partial_\nu A_\mu^{a(n)}(x)|^2+\frac{n^2}{R^2} |A_\mu^{a(n)}(x)|^2 
\biggr) \nonumber \\
&& +\sum_{n=1}^\infty \int \ d^4x \biggl({J_n^{a\mu}}^*(x)A_\mu^{a(n)}(x)
+J_n^{a\mu}(x){A_\mu^{a(n)}}^*(x)\biggr),
\label{ea3}
\eea
where we have used the notation 
\bea
J_n^{a\mu}(x)=\biggl(q^\dagger_L(x) {\bar \sigma}^\mu t^a q_L(x) e^{inL/2R}+
q^\dagger_R(x) \sigma^\mu t^a q_R(x)e^{-inL/2R}\biggr).
\label{j}
\eea
The dots in \eq{ea1} represent cubic and quartic interactions of the
gauge fields. These will not be relevant to the leading order analysis
in the weak coupling limit discussed below.

We have already discussed the integration of the massless gluon
degrees of freedom from the action $S_0$. Integrating out the massive
Kaluza-Klein modes from the action \eq{ea3} is a much simpler task. To
leading order in the gauge coupling, the effective four-fermi
interaction due to the exchange of these modes is given by
\bea
S_{1~{\rm eff}}=-g_4^2\sum_{n=1}^\infty 
\int d^4x \ d^4y \ \Delta_n(x-y) {J_{n\mu}^a}^*(x)J_n^{a\mu}(y),
\label{ea4}
\eea
where 
\bea
\Delta_n(x-y)=\int \frac{d^4k}{(2\pi)^4} \frac{e^{ik.(x-y)}}
{({\bar k}^2+\frac{n^2}{R^2})}, \qquad 
{\bar k}^2 \equiv {\vec k}^2-k_0^2-i\epsilon.
\label{g-1}
\eea
is the Feynman propagator for a scalar of mass $\frac{n}{R}$. Using
the Fierz identities (3.77) and (3.80) of \cite{PS} and, as before,
retaining only interaction terms between left and right-handed Weyl
components, the effective action \eq{ea4} becomes
\bea
S_{1~{\rm eff}}=2g_4^2 \int d^4x \ d^4y \biggl(\sum_{n=1}^\infty 
\cos(\frac{nL}{R}) \Delta_n(x-y)\biggr)[q^{\dagger\alpha}_L(x)q^\beta_R(y)]
[q^{\dagger\beta}_R(y)q^\alpha_L(x)].
\label{ea5}
\eea
Now, using the identity 1.445.2 of \cite{GR}, 
$$\sum_{n=1}^\infty \frac{\cos ns}{n^2+a^2}=\frac{\pi}{2a}
\frac{\cosh a(\pi-s)}{\sinh \pi a}-\frac{1}{2 a^2} \ ,$$
we get
\bea
S_{1~{\rm eff}}=g_4^2 \int d^4x \ d^4y \ G_1(x-y)
[q^{\dagger\alpha}_L(x)q^\beta_R(y)][q^{\dagger\beta}_R(y)q^\alpha_L(x)],
\label{ea6}
\eea
where
\bea
G_1(x)=\int \frac{d^4k}{(2\pi)^4} \ e^{ik.x} \ 
\tilde G_1(k), \qquad
\tilde G_1(k)=\frac{\pi R \cosh{\bar k}(\pi R-L)}
{{\bar k} \sinh{\bar k}\pi R}-\frac{1}{{\bar k}^2}.
\label{g1}
\eea

A few comments are in order: 
\begin{itemize}
\item The interaction in \eq{ea6}, \eq{g1} is exact for all values of 
$L$ and $R$, since we have summed over the exchange of all the
Kaluza-Klein modes. In particular, in the limit $R \rightarrow
\infty$, keeping $L$ fixed, we get
\bea
\tilde G_1(k) \rightarrow \frac{\pi R}{\bar k}e^{-{\bar k}L}.
\label{g1'}
\eea
\item For finite $R$, howsoever large, the second term in
$\tilde G_1(k)$ cancels the singularity in the first term in the limit
${\bar k} << 1/\pi R$. This is consistent with our expectation that
the range of the effective interaction \eq{ea6} should be of order the
Kaluza-Klein radius $R$. In fact, for fermion momenta much smaller
than $1/R$, one can approximate this non-local interaction by a local
NJL term, with small derivative corrections.
\item For ${\bar k} >> 1/L$, the second term on the right hand side in
\eq{g1} dominates, giving rise to a potentially problematic
short-distance interaction with a `wrong' sign. However, this term is
cancelled by the large $\bar k$ contribution to the total effective
action coming from the zero mode action, \eq{g01} and \eq{ea23}. The
net result is that for $\bar k >> 1/L$, $G_1(k)$ has the behaviour
given in \eq{g1'}. 
\end{itemize}

\subsection{\label{njlss}Non-local NJL from SS model}

Combining \eq{ea6} with \eq{ea24}, we get the total effective fermion
action
\bea
S_{\rm eff} &=& i\int \ d^4x \biggl(q^{\dagger\alpha}_L(x) 
{\bar \sigma}^\mu \partial_\mu q^\alpha_L(x)+
q^{\dagger\alpha}_R(x) \sigma^\mu \partial_\mu q^\alpha_R(x)\biggr)
\nonumber \\
&& +g_4^2 \int d^4x \ d^4y \ G(x-y)
[q^{\dagger\alpha}_L(x)q^\beta_R(y)][q^{\dagger\beta}_R(y)q^\alpha_L(x)],
\label{ea9} 
\eea
where
\bea
G(x) &=& \int \frac{d^4k}{(2\pi)^4} \ e^{ik.x} \  \tilde G(k),
\label{g}
\eea
and $\tilde G(k)=\tilde G_0(k)+\tilde G_1(k)$. 
Although a precise derivation of $\tilde G_0(k)$ does not exist, it
must satisfy certain conditions which we discussed in the previous
section. As a result of these, $\tilde G(k)$ must satisfy:
\begin{itemize}

\item For ${\bar k} \sim \Lambda$, $\tilde G(k) \sim 1/(k^2+\Lambda^2)$. 
This follows from the hierarchy of scales $\Lambda << 1/\pi R \leq
1/L$ and the fact that the range of the non-local quartic fermion
interaction in \eq{ea9} is set by the mass gap dynamically generated
in the $4$-dimensional Yang-Mills action \eq{ea2}, which is of order
the glueball mass $\sim \Lambda$.

\item For $\Lambda << {\bar k} << 1/\pi R$, $\tilde G(k) \sim
1/{\bar k}^2$. This is because for these values of $\bar k$, the
$4$-dimensional description continues to be valid and we are in the
asymptotically free regime.

\item For ${\bar k} >> 1/\pi R$, $\tilde G(k) \sim 
\pi R e^{-{\bar k}L}/{\bar k}$.
Here we have assumed $L << \pi R$ (which is automatically satisfied in
the limit of large $R$, keeping $L$ fixed). In this regime, the
$4$-dimensional description is inadequate since the Kaluza-Klein
states can be easily excited. We must now use the full $5$-dimensional
description. The exponential fall-off implies that the quartic fermion
interaction in \eq{ea9} has a short-distance cut-off as well. The
effective scale here is of order $L$, the separation between flavour
$D8$ and anti-$D8$-branes.

\end{itemize}
A simple function that contains all the three scales and seems to
capture all the essential features discussed above is
\bea
\tilde G(k)=\biggl(\frac{1+\pi R \bar k}{\bar k^2+\Lambda^2}\biggr)
e^{-{\bar k}L}.
\label{effg}
\eea
This differs from \eq{g03} in that $1/L$ enters as a smooth
ultraviolet scale, as opposed to the hard cut-off $M$ that came with
\eq{g03}. Moreover, it contains the additional scale $R$, which has to
do with the underlying $5$-dimensional origin of the model. We will
use this simpler function in the analysis that follows. Also,
throughout the following we will assume $L << \pi R$.

\section{\label{chisb}$\chi$SB in non-local NJL model}

As usual, we first introduce scalars to rewrite \eq{ea9} in the
equivalent form
\bea
S_{\rm eff} &=& i\int \ d^4x \biggl(q^{\dagger\alpha}_L(x) 
{\bar \sigma}^\mu \partial_\mu q^\alpha_L(x))+
q^{\dagger\alpha}_R(x) \sigma^\mu \partial_\mu q^\alpha_R(x)\biggr)
\nonumber \\
&& +\int d^4x \int d^4y \biggl[-\frac{{T^{\alpha \beta}}^*(x,y)
T^{\alpha \beta}(x,y)}{g_4^2 G(x-y)}+{T^{\alpha \beta}}^*(x,y)
q^{\dagger\beta}_R(y)q^\alpha_L(x) \nonumber \\
&& \hspace{4cm} +T^{\alpha \beta}(x,y)
q^{\dagger\alpha}_L(x)q^\beta_R(y)\biggr].
\label{tact1}
\eea
It is easy to verify the equivalence of this action to \eq{ea9} by
using the equations of motion of the scalars,
\bea
T^{\alpha \beta}(x,y)=g_4^2 \ G(x-y) \ q^{\dagger\beta}_R(y)q^\alpha_L(x).
\label{t1}
\eea 
The next step is to integrate out the fermions to get an effective
action for the scalars. In the large-$N_c$ limit, a classical
treatment is adequate. Since we are only interested in the solution
corresponding to the ground state, which is Poincare invariant and
invariant under diagonal (vector) flavour group, we may use the ansatz
$T^{\alpha\beta}(x,y)=\delta^{\alpha \beta} T(|x-y|)$. This simplifies
calculation of the effective action. Making a Wick rotation to the
Euclidean signature, we get
\bea
\frac{S^E_{\rm eff}}{VN_cN_f}=\frac{1}{g_4^2N_c}\int d^4x \ 
\frac{|T(x)|^2}{G^E(x)}-\int \frac{d^4k}{(2\pi)^4} \ 
{\rm ln}\biggl(1+\frac{|\tilde T(k)|^2}{k^2}\biggr),
\label{tact2}
\eea
where $V$ is the $4$-volume and $\tilde T(k)$ is related to $T(x)$ by
a Fourier transform, which we define by the relation \eq{g02} for any
function. Also, $G^E(x)$ is the Euclidean version of $G(x)$. Taking
the Fourier transform of \eq{effg}, we get
\bea
G^E(x)=\frac{1}{4\pi^2|x|}\int_0^\infty dk \ k^2 J_1(k|x|)
\biggl(\frac{1+\pi R k}{k^2+\Lambda^2}\biggr)e^{-kL},
\label{g04}
\eea
where $k$ is the magnitude of the Euclidean four-momentum and
$J_1(k|x|)$ is a standard Bessel function. The $k$-integral can be done
using the identity
\bea
\frac{1+\pi R k}{k^2+\Lambda^2}=\frac{1+i\pi R \Lambda}{2i\Lambda}
\int_0^\infty ds \ e^{is\Lambda} \ e^{-sk}+c.c.,
\label{iden}
\eea
and the identity 6.623.2 of \cite{GR}. After some simplification, the
final expression takes the form
\bea
G^E(x)=\frac{\Lambda^2}{4\pi^2} \ g(|x|\Lambda),
\label{g040}
\eea
where
\bea
g(r)=\frac{R_\Lambda}{(L_\Lambda^2+r^2)^{3/2}} 
+(\cos L_\Lambda+R_\Lambda \sin L_\Lambda) \ \mathcal{I}_1(r)  
+(\sin L_\Lambda-R_\Lambda \cos L_\Lambda) \ \mathcal{I}_2(r). 
\label{g041}
\eea
In the above, we have introduced the dimensionless quantities
\bea
R_\Lambda=\pi R \Lambda, \qquad L_\Lambda=L \Lambda.
\label{defII}
\eea
Moreover,
\bea
\mathcal{I}_1(r) &=& \int^\infty_{L_\Lambda} ds \frac{\cos s}{(s^2+r^2)^{3/2}}
=\frac{K_1(r)}{r}-\int^{L_\Lambda}_0 ds \frac{\cos s}{(s^2+r^2)^{3/2}}, 
\nonumber \\
\mathcal{I}_2(r) &=& \int^\infty_{L_\Lambda} ds \frac{\sin s}{(s^2+r^2)^{3/2}}
=\frac{1}{r}-\frac{\pi}{2r}(I_1(r)-\mathbf{L}_1(r))-
\int^{L_\Lambda}_0 ds \frac{\sin s}{(s^2+r^2)^{3/2}},
\label{defI}
\eea
where $K_1(r)$ and $I_1(r)$ are standard Bessel functions and
$\mathbf{L}_1(r)$ is a Struve function. It turns out that numerical
calculations are done faster with the second form of
$\mathcal{I}_{1,2}(r)$. Note that in the region $r << R_\Lambda$, all
the $5$ dimensions must come into play. Indeed, in this region we get
\bea
g(r) \approx \left\{ \begin{array}{rcl}
R_\Lambda/L_\Lambda^3 & \mbox{for}
& r << L_\Lambda \\ R_\Lambda/r^3 & 
\mbox{for} & L_\Lambda << r << R_\Lambda  \end{array}\right.
\label{g42}
\eea
For $r >> R_\Lambda$, the system becomes effectively
$4$-dimensional. Here we find 
\bea
g(r) \approx  \left\{ \begin{array}{rcl}
1/r^2 & \mbox{for}
& 1 >> r >> R_\Lambda \\ \sqrt{\frac{\pi}{2}} e^{-r}/r^{3/2} & 
\mbox{for} & r >> 1  \end{array}\right.
\label{g43}
\eea
\begin{figure}[htb] 
\centering 
\includegraphics[height=6cm,width=9cm]{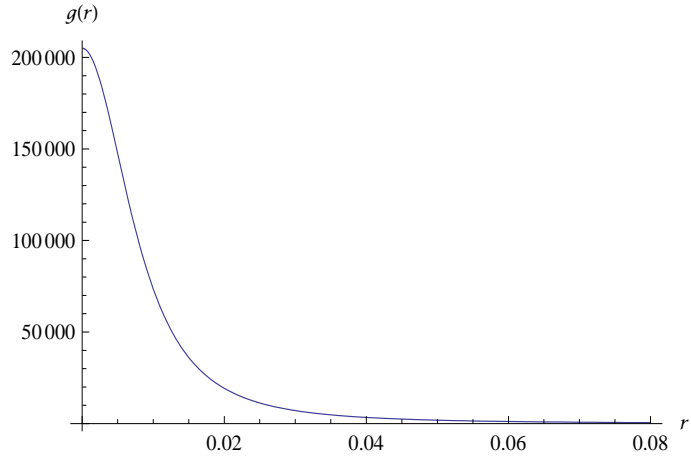}
\parbox{5in}{\caption{$g(r)$ as a function of $r$ for the 
parameter values $L_\Lambda=0.01$ and $R_\Lambda=0.2$.}
\label{Fig.1}}
\end{figure}
In \fig{Fig.1} we have plotted the function $g(r)$ as a function of
$r$. We see that it approaches a constant for $r << L_\Lambda$. For $r
\gtrsim L_\Lambda$, it decreases very rapidly and eventually for
$r> 1$ (far beyond the region shown in the figure) it decays
exponentially \footnote{This behaviour
actually holds only for intermediate values of $r$ which satisfy
$r^{7/2}e^{-r} > R_\Lambda$. If $R_\Lambda$ is small, this inequality
can allow rather large values of $r$. At much larger values of $r$,
$g(r)$ decays only as $1/r^5$. This is presumably an artifact of the
choice of the $\tilde G(k)$ function we have made in \eq{effg}. For
practical reasons, in the numerical calculations done in the next
section, we have simply set $g(r)$ to zero beyond a sufficiently large
value of $r$. This is consistent with the expectation that the
interactions should decay exponentially beyond the confinement scale,
$r=1$.}.

\subsection{\label{geop}Gap equation and order parameter of $\chi$SB}

The equation of motion for $T(x)$ following from the effective action
\eq{tact2} is
\bea
\frac{1}{g_4^2N_c}\int d^4x \ \frac{T(x)}{G^E(x)} e^{-ik.x}
=\frac{\tilde T(k)}{k^2+|\tilde T(k)|^2}
\label{teq1}
\eea 
This nonlinear equation for the order parameter is the analogue of the
gap equation for the present case. $T(x)=0$ is the trivial solution
which preserves chiral symmetry. However, this is not the solution
which minimizes the effective action \eq{tact2}. To see this
\cite{AHJK}, multiply \eq{teq1} on both sides by $\tilde T^*(k)$ and
integrate over $k$. This gives
\bea
\frac{1}{g_4^2N_c}\int d^4x \ \frac{|T(x)|^2}{G^E(x)}=
\int \frac{d^4k}{(2\pi)^4}\frac{|\tilde T(k)|^2}{k^2+|\tilde T(k)|^2}.
\label{energy}
\eea
Using this in \eq{tact2}, we get
\bea
\frac{S^E_{\rm eff}}{VN_cN_f}=\int \frac{d^4k}{(2\pi)^4} 
\biggl[\frac{|\tilde T(k)|^2}{k^2+|\tilde T(k)|^2}-{\rm
ln}\biggl(1+\frac{|\tilde T(k)|^2}{k^2}\biggr)\biggr],
\label{tact3}
\eea
It is easy to see that the integrand on the right-hand side above is a
decreasing function of $|\tilde T(k)|/k$ and that it vanishes for
$\tilde T(k)=0$. It follows that $\tilde T(k)=0$ is not the solution
which minimizes the effective action
\eq{tact2}. We also note that a potential divergence from the large
$k$ end gets cancelled between the two terms in the integrand and the
net result of the integration over $k$ is finite, provided $\tilde
T(k)$ is a decreasing function for large $k$. If such a solution
exists, then the chiral symmetry is spontaneously broken.

The order parameter of chiral symmetry breaking is the condensate
\bea
\phi(x)=\frac{1}{N_c} \langle q^{\dagger\alpha}_L(x)q^\alpha_R(0) \rangle.
\label{op1}
\eea 
The field $T(x)$ is related to it by \eq{t1}, i.e.
\bea
T(x)=4\pi^2\lambda \ G^E(x) \ \phi(x), \qquad \lambda \equiv g_4^2 N_c/4\pi^2.
\label{t2}
\eea
The gap equation \eq{teq1} can be rewritten in terms of $\phi(x)$ as
\bea
\tilde\phi(k)=\frac{\tilde T(k)}{k^2+|\tilde T(k)|^2}.
\label{teq2}
\eea 
We will look for solutions of this equation with $\phi(x)$, and hence
$T(x)$, real. Since these are spherically symmetric functions of
$|x|$, their Fourier transforms, $\tilde \phi(k)$ and $\tilde T(k)$,
are also real functions of $k$. Furthermore, we see from \eq{teq2}
that for large $k$, $\tilde\phi(k) << 1/k$. This is because $\tilde
T(k)$ must be a decreasing function for large $k$, for reasons
explained above. Now, solving \eq{teq2} for $\tilde T(k)$ as a function
of $\tilde\phi(k)$, we get
\bea
\tilde T_\pm(k)=\frac{1}{2\tilde\phi(k)}
\biggl[1 \pm \sqrt{1-4k^2\tilde\phi^2(k)}\biggr].
\label{sol1}
\eea
This has real solutions only for $\tilde\phi(k) \leq 1/2k$. For large
$k$, with $\tilde\phi(k) << 1/k$, we get $\tilde T_+(k) >> k$ and
$\tilde T_-(k) << k$. So, the desired solution $\tilde T(k)$ must
coincide with $\tilde T_-(k)$ for large $k$. For small enough $k$, the
solution $\tilde T(k)$ must go to a constant (the mass gap) and so it
must coincide with $\tilde T_+(k)$ for small $k$. The transition from
one to the other occurs at some scale $k=\mu$, where the two solutions
coincide, i.e. $\tilde T_+(\mu)=\tilde T_-(\mu)$. From \eq{sol1} we
see that $\mu$ satisfies the equation $\tilde\phi(\mu)= 1/2\mu$.

\subsection{\label{solge}Solutions of the gap equation}

We parametrise the condensate as follows:
\bea
\phi(x)=\frac{\phi_0}{4\pi^2 l^3} \varphi(|x|/l).
\label{paramtz}
\eea
Here $l$ is the $\chi$SB scale. The normalization has been chosen to
explicitly display the dimensions of the order parameter and to have
$\varphi(0)=1$. Now, using \eq{t2}, the Fourier transforms, $\tilde
\phi(k)$ and $\tilde T(k)$, can be written as
\bea
&& \tilde\phi(k)=l \phi_0 \ f(kl), \qquad \quad~
f(p) \equiv \frac{1}{p}\int^\infty_0 dy \ y^2 J_1(py) \ \varphi(y), 
\nonumber \\
&& \tilde T(k)=\lambda l_\Lambda \Lambda \phi_0 \ t(kl), \qquad
t(p) \equiv \frac{1}{p}\int^\infty_0 dy \ y^2 J_1(p y) \ g(l_\Lambda y) \ 
\varphi(y), 
\label{fts}
\eea
where $l_\Lambda \equiv l\Lambda$ and $J_1$ is a standard Bessel
function. Using these, we can rewrite the gap equation \eq{teq2} as
\bea
f(p)=\frac{\bar\lambda t(p)}
{p^2+\bar\lambda^2 \phi_0^2 t^2(p)}, \qquad 
\bar\lambda \equiv \lambda l_\Lambda.
\label{teq3}
\eea
This is a nonlinear equation which we are unable to solve
analytically. However, there are some general observations we
can make.
\begin{itemize}
\item\noindent Equation \eq{teq3} cannot have a solution with the
$\chi$SB scale $l$ arbitrarily smaller than $L$. On physical grounds,
we expect $\varphi(|x|/l)$ to be substantially different from zero
only in the region $0 \leq |x| \lesssim l$, vanishing rapidly for
$|x| >> l$. As a result, most of the contribution to the integral in
the definition of $t(p)$ in \eq{fts} comes from a region in which
the argument of the function $g(r)$ varies over the range $[0,l_\Lambda]$. For
$l << L$, $g(r)$ is roughly a constant
($=\frac{R_\Lambda}{L_\Lambda^3}$) over this range of $r$, as can be
seen from \eq{g041}. So, for $l << L$, $t(p) \approx
\frac{R_\Lambda}{L_\Lambda^3} f(p)$. But this is not a solution to
\eq{teq3}, as can be easily checked. This argument works even better
for large $p$, because then the contribution to the integral from $y >
1$ region is even more suppressed, beyond that due to a rapidly
falling $\varphi(y)$. But, for large $p$ a solution to
\eq{teq3} must satisfy $t(p) \propto p^2 f(p)$. Thus, the gap equation
\eq{teq3} has no solutions for $l << L$. 

\item\noindent In principle, $\chi$SB solutions to the gap equation
should exist for all $l \geq L$. This is because our non-local NJL
model does not incorporate confinement. However, for solutions which
have $l > \Lambda^{-1}$, i.e. $l_\Lambda > 1$, it should be possible
to replace the non-local model by an effective local NJL model with
$\Lambda$ as the ultraviolet cut-off. This is because of the
exponential decay of the four-fermi interaction $g(r)$ for $r >
1$. Since the local NJL model has no $\chi$SB solutions below a
critical coupling, we expect a critical coupling to show up for values
of order $l_\Lambda \gtrsim 1$ in the present non-local NJL model as
well. One way this can happen is that as $l_\Lambda$ increases beyond
$L_\Lambda$, the value of $\lambda$ which gives rise to this solution
decreases until it hits a critical value at around $l_\Lambda \sim
1$. This argument cannot be made for the non-compact SS model
considered in \cite{AHJK} because of the absence of the mass scale
$\Lambda$ and the consequent absence of the exponential decay of
$g(r)$ for $r > 1$.
\end{itemize} 

Numerical calculations reported below bear out both the above
expectations.

\subsection{\label{numerics}Numerical solutions}

In the following, we will report on some solutions to the gap equation
\eq{teq3} obtained numerically using mathematica. This numerical work
is based on the following strategy. From the expected physical
properties of the order parameter $\phi(x)$, we first make an ansatz
for it:
\bea
\varphi(x)=\frac{e^{-x}}{(c^2x^2+1)^\sigma}.
\label{ansatz}
\eea
The power $\sigma$ and the constant $c$ are adjustable
parameters. With the $\chi$SB scale $l$ and the normalization constant
$\phi_0$, there are altogether four adjustable parameters. Given
\eq{ansatz}, the two sides of \eq{teq3} can be computed and compared, 
and the difference can be minimized by varying these parameters. The
numerical computations were done as follows. For the left-hand side of
\eq{teq3}, we need to calculate $f(p)$. This can be done once we
choose some values for $\sigma$ and $c$. After some experience with
the calculations, it was not hard to make a good guess for the right
values. For the right-hand side of
\eq{teq3}, we need $t(p)$ for which one first needs to calculate the
function $g(r)$. This requires choosing values for $R_\Lambda$ and
$L_\Lambda$. In \fig{Fig.1} we have shown an example of $g(r)$ for
$R_\Lambda=0.2$ and $L_\Lambda=0.01$. With $g(r)$ at hand, one can now
calculate $t(p)$, after making a choice for $l_\Lambda$. The
right-hand-side of \eq{teq3}, which we denote as $f_t(p)$,
\bea
\frac{\bar\lambda t(p)}
{p^2+\bar\lambda^2 \phi_0^2 t^2(p)}
\equiv f_t(p),
\label{ftp}
\eea
can then be computed. This requires making a choice for the parameters
$\bar\lambda$ and $\phi_0$, which are adjusted such that the deviation
$\int dp (f(p)-f_t(p))^2$ is minimized. In principle, in the
calculation of the deviation, the range of the integral over $p$
should extend to infinity. In practice, we have found a value of about
$10$ to be good enough for the upper limit (for the values of the
parameters $L_\Lambda$ and $R_\Lambda$ we have used in our
calculations), in the sense that the value of the integral remains
essentially unchanged if the upper limit is increased beyond this
value. The whole procedure was then repeated with slightly different
values of $\sigma$, $c$, $\bar\lambda$ and $\phi_0$ until the
deviation was minimized for the chosen values of $R_\Lambda$,
$L_\Lambda$ and $l_\Lambda$. The value of $\lambda$ relevant to these
values of the parameters was obtained from its relation to
$\bar\lambda$ given in \eq{teq3}.

\begin{figure}[htb] 
\centering 
\begin{minipage}[b]{6.5cm}
    \includegraphics[height=4.5cm,width=6cm]{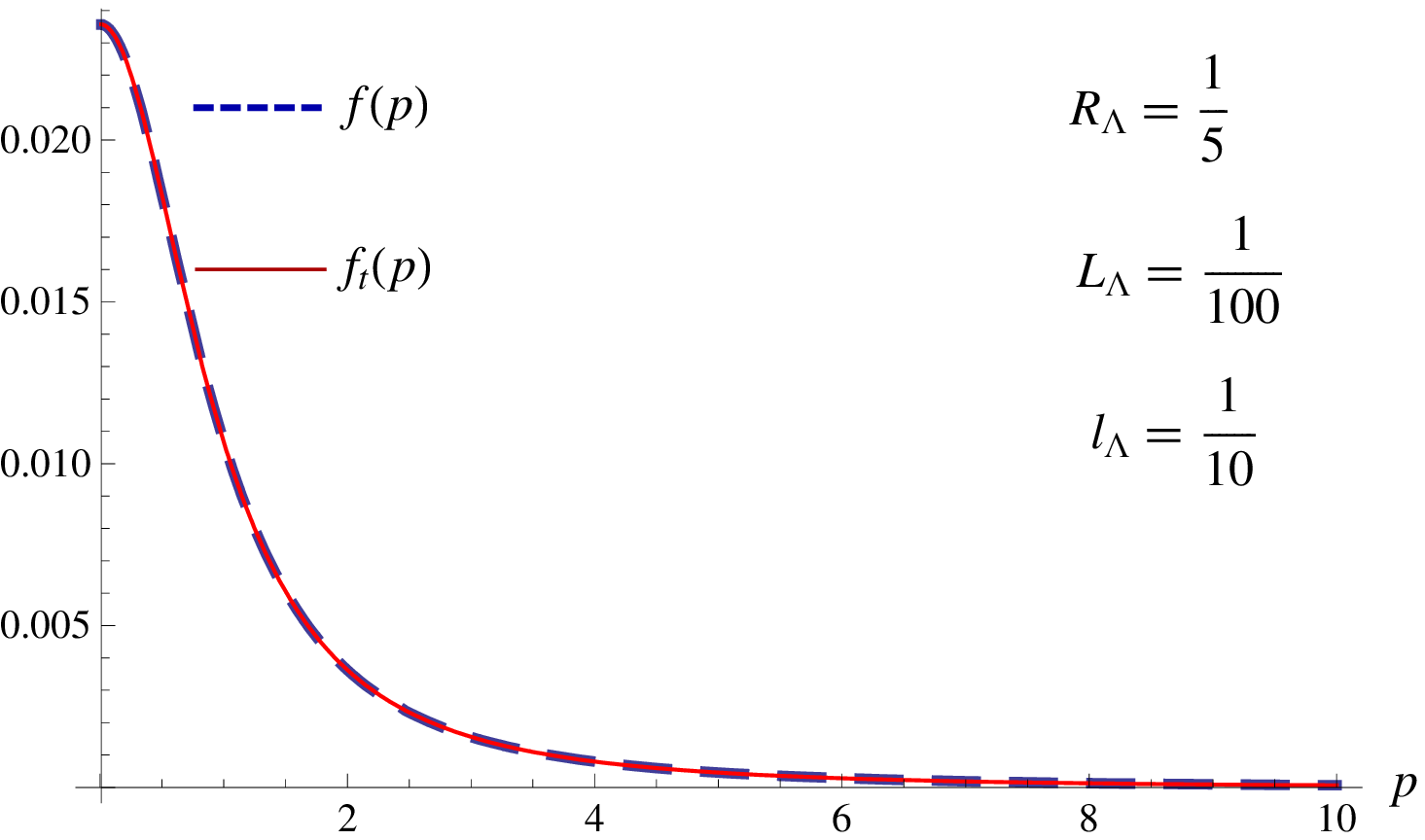}  
  \end{minipage}
  \begin{minipage}[b]{6.5cm}
    \includegraphics[height=4.5cm,width=6cm]{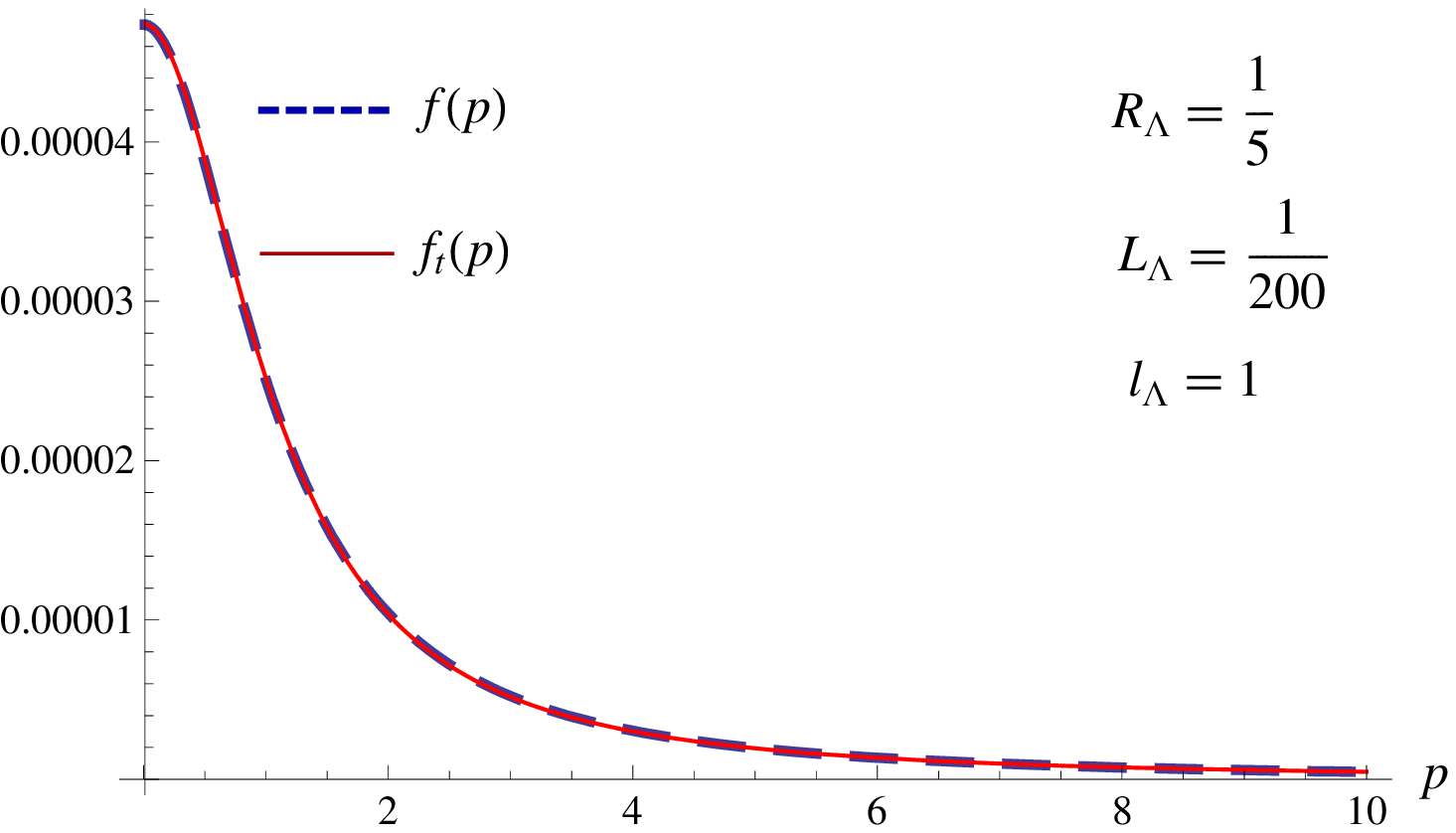} 
  \end{minipage}
\parbox{5in}{\caption{The functions $f(p)$ and $f_t(p)$ shown in these figures 
are the two sides of the gap equation. The two figures correspond 
to two different sets of values of $\{R_\Lambda,L_\Lambda\}$, as 
indicated.}
\label{fig.23}}
\end{figure}
In \fig{fig.23} we have given two examples of the quality of solutions
obtained in this way for two different sets of values of $\{R_\Lambda,
L_\Lambda\}$, with $l_\Lambda >> L_\Lambda$. The agreement between
$f(p)$ and $f_t(p)$ is excellent.  In fact, the deviation $\int dp
(f(p)-f_t(p))^2$ is less than a thousandth of a percent of the quantity
$\int dp (f(p))^2$. Surprisingly, in all the solutions that we have
obtained, the value of the parameter $c$ which minimizes the deviation
turns out to be exactly $l/L$. This may provide a hint for analytical
solutions of the gap equation.

Note that both cases in \fig{fig.23} have $l_\Lambda >> L_\Lambda$. As
argued in the previous subsection, we do not expect any solution for
$l_\Lambda << L_\Lambda$. In fact, our numerical calculations indicate
that there is no solution for $l_\Lambda \lesssim L_\Lambda$.
\begin{figure}[htb] 
\centering 
\includegraphics[height=4.5cm,width=6cm]{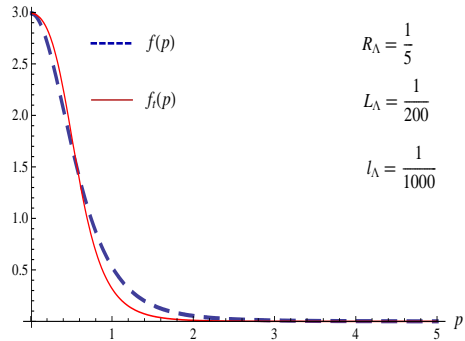}
\parbox{5in}{\caption{The two sides of the gap equation for 
$l_\Lambda=L_\Lambda/5$.}
\label{fig.4}}
\end{figure}
This can be seen from the example given in \fig{fig.4} where we have
taken $l_\Lambda= L_\Lambda/5$. There is no agreement, which is the
best we have been able to do with the ansatz \eq{ansatz}. The seeming
agreement at large $p$ is misleading because the values of both $f(p)$
and $f_t(p)$ are so small that the figure cannot distinguish them from
zero at the scale used. 
\begin{figure}[htb] 
\centering 
\begin{minipage}[b]{6.5cm}
    \includegraphics[height=4.5cm,width=6cm]{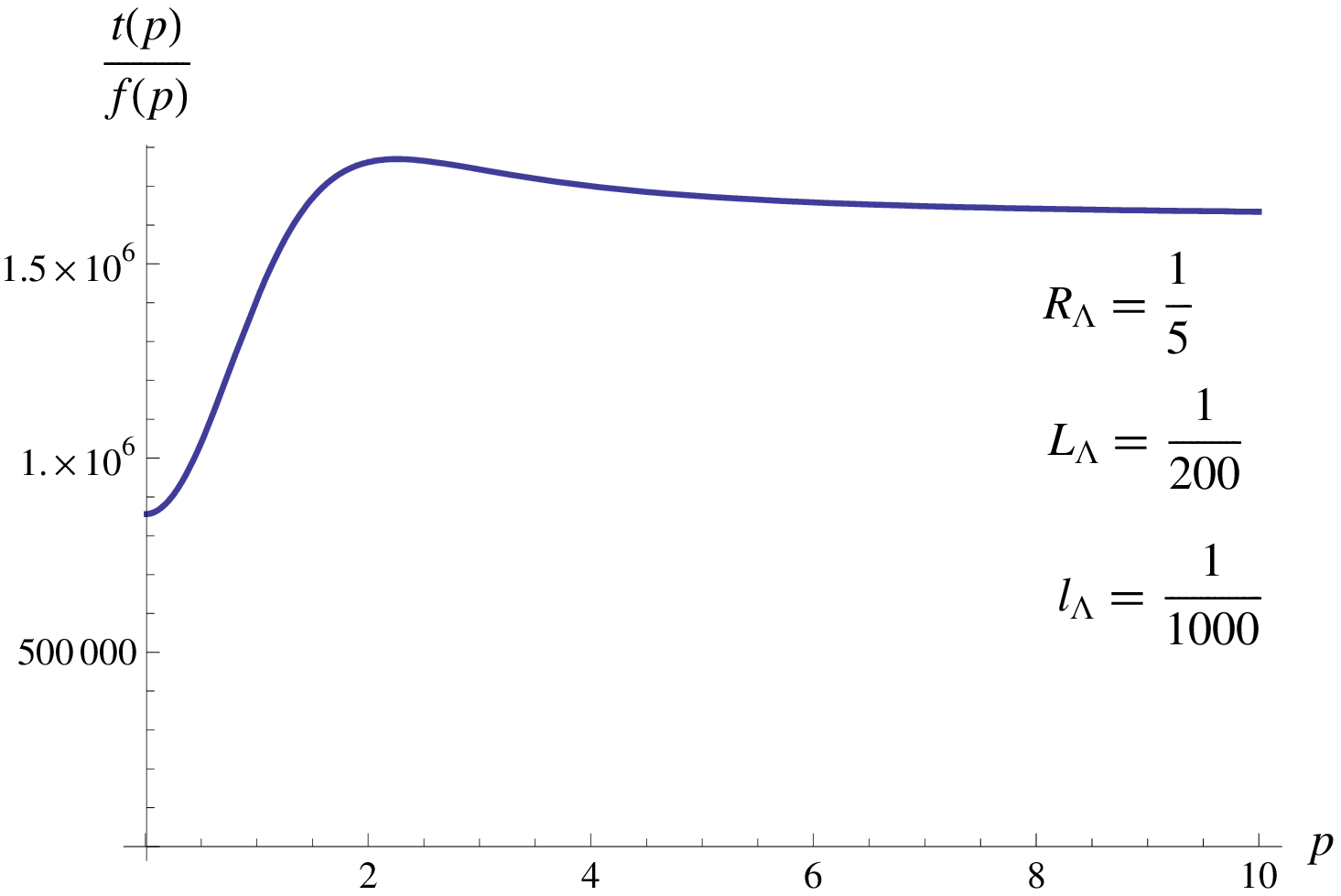}  
  \end{minipage}
  \begin{minipage}[b]{6.5cm}
    \includegraphics[height=4.5cm,width=6cm]{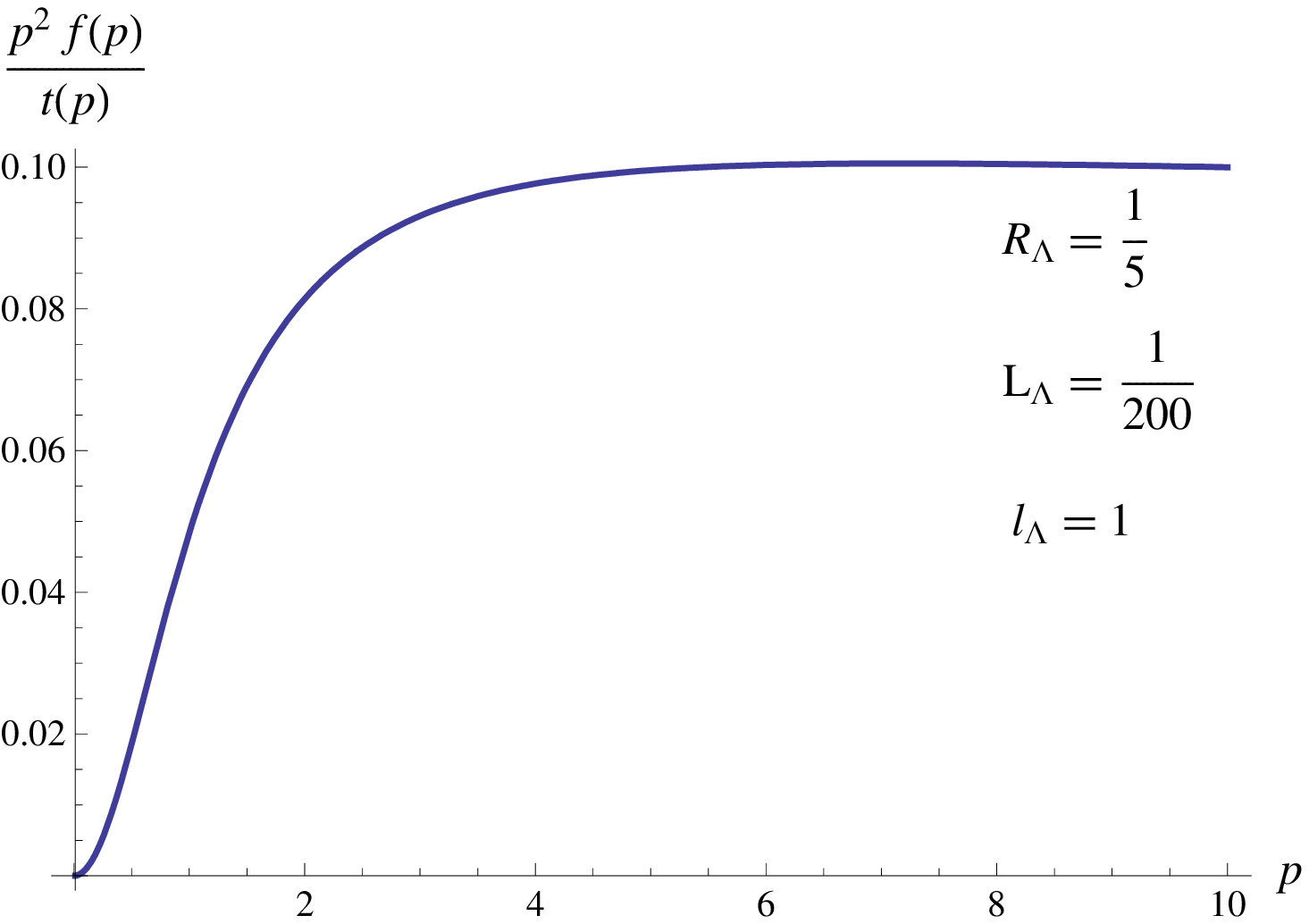}
  \end{minipage}
\parbox{5in}{\caption{The first figure shows $t(p)/f(p)$ for 
$l_\Lambda < L_\Lambda$
and the second figure shows $p^2 f(p)/t(p)$ for $l_\Lambda >>
L_\Lambda$ as a function of $p$.}
\label{fig.56}}
\end{figure}
A better measure for the behaviour at large
$p$ is the ratio $t(p)/f(p)$, which is expected to approach a constant
at large $p$ for $l_\Lambda << L_\Lambda$. In the first of
\fig{fig.56} we have plotted this ratio. It behaves as predicted at
large $p$. The value of the constant also turns out to be close to the
expected one, namely $R_\Lambda/L_\Lambda^3$. For comparison, in the
second of \fig{fig.56} we have plotted the ratio $p^2 f(p)/t(p)$ for
the second example of \fig{fig.23}, which is expected to approach a
constant at large $p$ since this provides a solution to the gap
equation. The figure verifies this, implying that in this case
$t(p) \sim p^2 f(p)$ at large $p$.

\begin{figure}[htb] 
\centering 
\begin{minipage}[b]{6.5cm}
    \includegraphics[height=4.5cm,width=6cm]{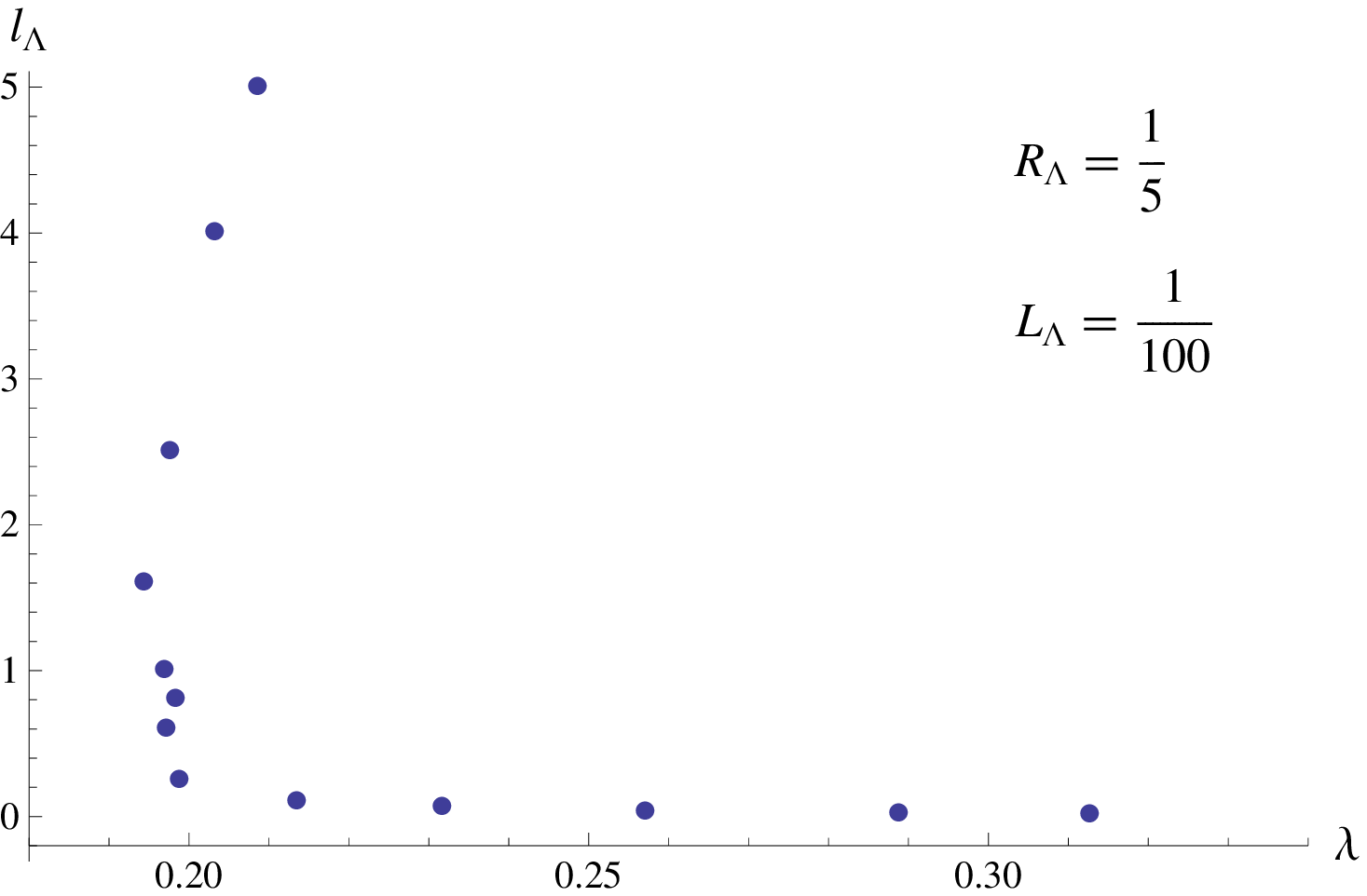}  
  \end{minipage}
  \begin{minipage}[b]{6.5cm}
    \includegraphics[height=4.5cm,width=6cm]{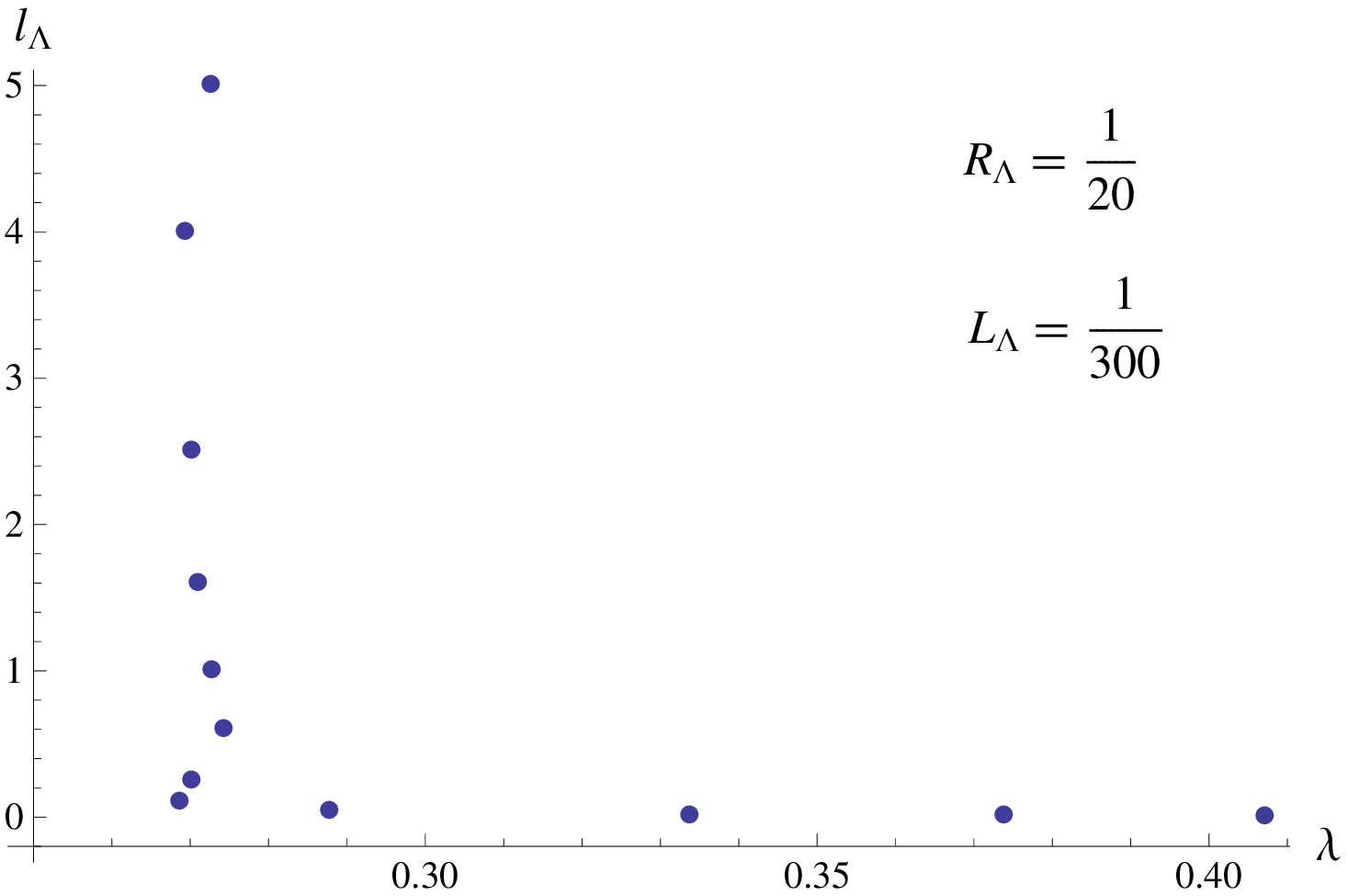}
  \end{minipage}
\parbox{5in}{\caption{$l_\Lambda$ as a function of $\lambda$ for 
two different sets 
of values of $\{R_\Lambda,L_\Lambda\}$.}
\label{fig.78}}
\end{figure}

In \fig{fig.78} we have plotted $l_\Lambda$ as a function of
$\lambda$. Figures for two different sets of values of
$\{R_\Lambda,L_\Lambda\}$ have been given.
Starting with $l_\Lambda=L_\Lambda$, we see that at first $\lambda$
decreases with increasing $l_\Lambda$, until it reaches a minimum at
around $l_\Lambda=1$. Beyond this point, increasing $l_\Lambda$ seems
to be accompanied by an unchanged or perhaps even an increasing
$\lambda$ \footnote{This point is difficult to clarify with much
accuracy beyond the range of values of $l_\Lambda$ shown in the
figure. This is because calculations for such large values of
$l_\Lambda$ require a greater precision in calculating $g(r)$ and
hence take much longer time.}. We have verified the behaviour shown in
the two examples in \fig{fig.78} for several other values of the set
$\{R_\Lambda,L_\Lambda\}$ and believe that this is the general
behaviour. These data provide fairly convincing evidence for the
existence of a critical value of $\lambda$ below which no solutions to
the gap equation exist.

\section{\label{noncompact}The non-compact limit}

It is of interest to ask what happens in the non-compact
limit, $R \rightarrow \infty$ keeping $L$ fixed. In our model, taking
this limit is somewhat subtle, because the hierarchy of scales $L <<
\pi R << \Lambda^{-1}$ must be maintained as the limit is taken. 
Furthermore, even though the non-local NJL model does not incorporate
confinement, to connect with the underlying weakly coupled gauge
theory we may wish to impose on this model the relations between the
confinement scale $\Lambda^{-1}$, $R$ and the coupling $g_5^2$, given
by
\eq{scale} and
\eq{YMg} (with $m \sim 1/\pi R$). Under the scaling $R \rightarrow R(\eta)
=\eta R$, the following scaling properties can be deduced from these
relations:
\bea
R_\Lambda(\eta)=(R_\Lambda)^{\eta}, \qquad L_\Lambda(\eta)=
L_\Lambda \frac{(R_\Lambda)^{\eta -1}}{\eta},
\label{scaling}
\eea 
where $L_\Lambda$ and $R_\Lambda$ are the values for $\eta=1$. The
non-compact limit corresponds to taking $\eta \rightarrow
\infty$. Since $R_\Lambda << 1$, this means that both $R_\Lambda(\eta)$ 
and $L_\Lambda(\eta)$ vanish in this limit.

\begin{figure}[htb] 
\centering 
   \begin{minipage}[b]{6.5cm}
    \includegraphics[height=4.5cm,width=6cm]{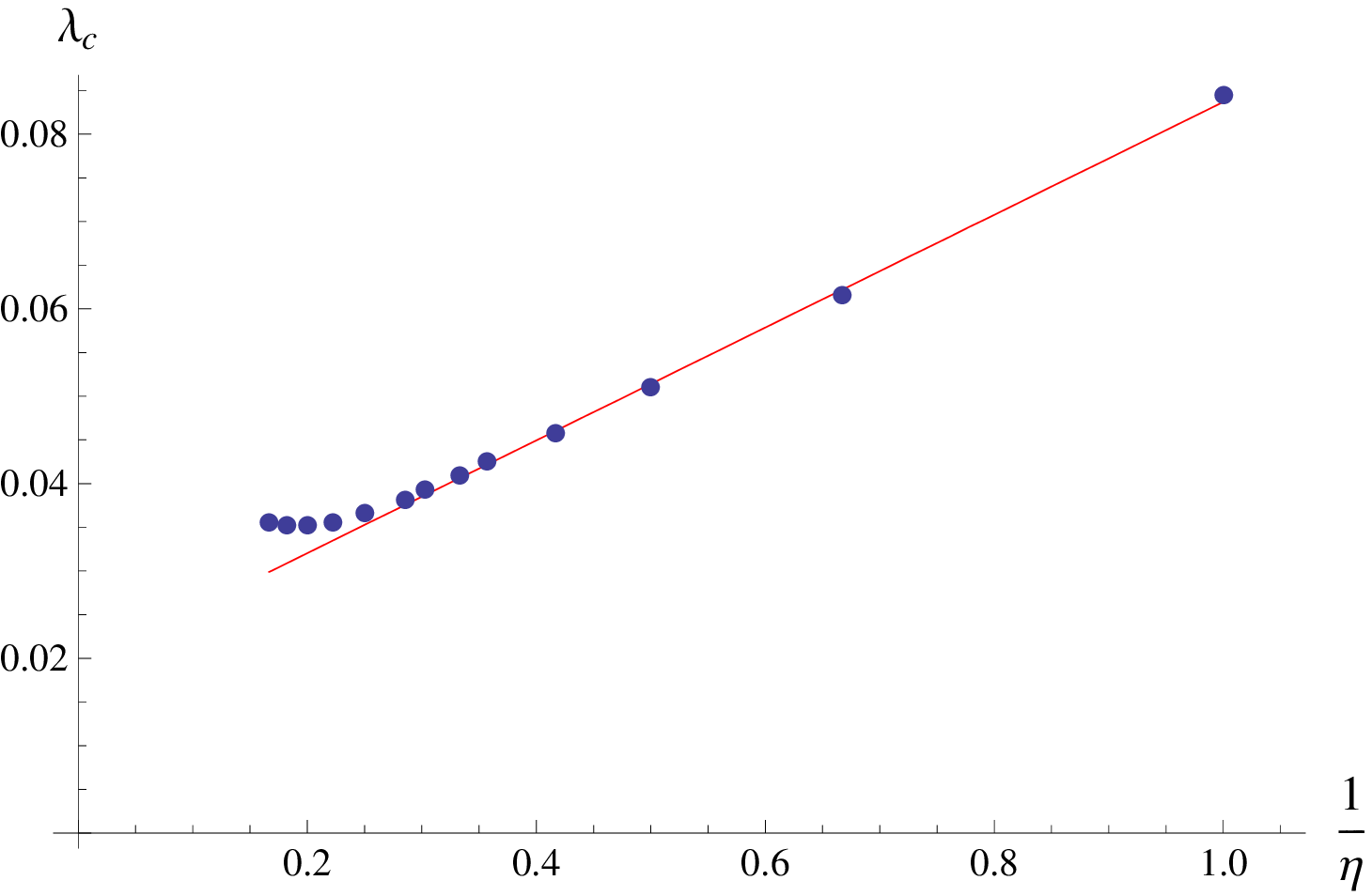}  
  \end{minipage}
  \begin{minipage}[b]{6.5cm}
    \includegraphics[height=4.5cm,width=6cm]{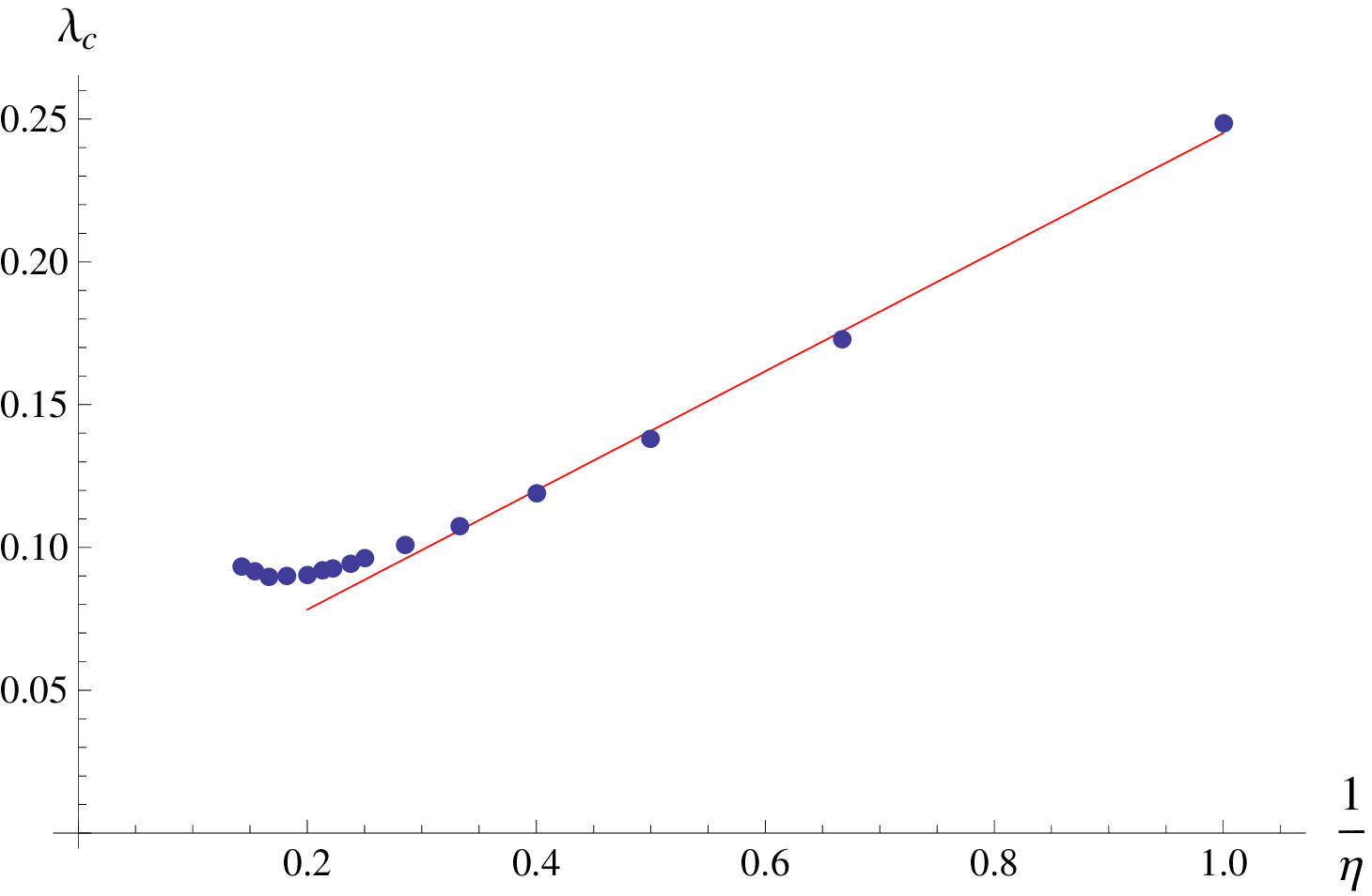}
  \end{minipage}  
\parbox{5in}{\caption{$\lambda_c$ as a function of $1/\eta$. 
The data in the two 
figures correspond to the scaling rule \eq{scaling} for two different
initial values of the set $\{R_\Lambda,L_\Lambda\}$, $\{1/2,1/100\}$
for the first figure and $\{1/2,1/30\}$ for the second. The solid
lines are drawn to indicate the region in which linear behaviour with
$1/\eta$ is seen. The general behaviour is similar in the two cases.}
\label{fig.910}}
\end{figure}

Let us denote by $\lambda_c$ the value of $\lambda$ for $l_\Lambda=1$.
$\lambda_c$ is close to the minimum value of $\lambda$ (see
\fig{fig.78}) and so may be considered to be the value of the critical
coupling. How does $\lambda_c$ change as a function of $\eta$? To find
this, we have numerically calculated $\lambda_c(\eta)$ for the set
$\{R_\Lambda(\eta),L_\Lambda(\eta)\}$ for different values of
$\eta$. The numerical data have been plotted in \fig{fig.910} as a
function of $1/\eta$. Calculations were done for two different sets of
values of $\{R_\Lambda,L_\Lambda\}$ to check dependence on initial
conditions. Numerical data in the two graphs of \fig{fig.910} show a
similar pattern, indicating no dependence of the general behaviour on
initial conditions.
The data show that at first, as $\eta$ grows away from the initial
value $\eta=1$, $\lambda_c$ decreases linearly with $1/\eta$, as one
might expect from the relation between the $4$-dimensional 't Hooft
coupling and its $5$-dimensional counterpart, $\lambda \equiv
\lambda_5/2\pi R$, which was used in deriving the scaling relations
\eq{scaling}. However, at some point, $\lambda$ stops falling and
seems to bottom out and start increasing or perhaps go to a constant
\footnote{More detailed calculations are needed to settle between these 
two possibilities. Calculations at higher values of $\eta$ involve
fine-tuning of various parameters of the solution to the gap equation
and so they are harder to determine. For this reason we have
restricted ourselves to a maximum of $\eta=5.5$.  It should be
possible to go to larger values with some more effort.}. This is not
consistent with the relation between the couplings, so it seems that
the non-compact limit cannot be reached by the one-parameter scaling
given by \eq{scaling}. 

An alternative way of taking the non-compact limit is as follows.  The
scaling rule \eq{scaling} used in the above analysis was derived from
relations between $\Lambda$, $R$, $\lambda$ and $\lambda_5$ which
follow from confinement in the underlying low-energy gauge-theory.
\begin{figure}[htb] 
\centering 
\includegraphics[height=4.5cm,width=6cm]{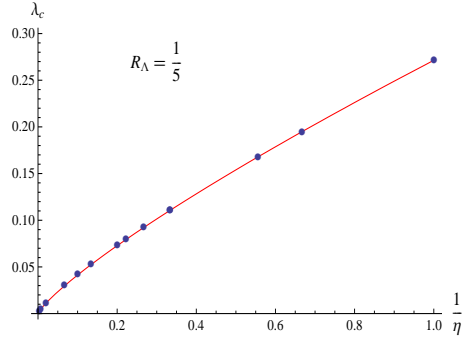}
\parbox{5in}{\caption{$\lambda_c$ as a function of $1/\eta$. These data
have been obtained for $L_\Lambda(\eta)=L_\Lambda/\eta$, with
$L_\Lambda=3/200$ and keeping $R_\Lambda$ fixed as $\eta$ is increased
from $1$. The solid line is a power-law fit to the data.}
\label{fig.11}}
\end{figure}
Since the non-local NJL model does not incorporate confinement, we may
wish to relax these conditions among the parameters. However, we must
still maintain the hierarchy of scales $L << \pi R <<
\Lambda^{-1}$. One simple way to do this is to keep $R_\Lambda$ fixed
(and $<< 1$) as $R$ is scaled. Under the scaling $R \rightarrow \eta
R$, then, we must have $\Lambda \rightarrow
\Lambda/\eta$. This implies that $L_\Lambda$ scales to $L_\Lambda/\eta$.

In \fig{fig.11} we have shown data for $\lambda_c(\eta)$ obtained as a
function of $1/\eta$ by computing the critical coupling for the set
$\{R_\Lambda, L_\Lambda(\eta)\}$ for different values of $\eta$.  The
data fit almost perfectly to a power law,
$\lambda_c=0.271339/\eta^{0.817788}$. Since this fit implies that $\eta
\lambda_c(\eta)$ blows up in the limit $\eta \rightarrow \infty$, the 
non-compact limit cannot be reached by this scaling either.

Our tentative conclusion is that possible $\chi$SB solutions in the
non-compact version of the non-local NJL model cannot be obtained by
taking any simple $R \rightarrow \infty$ limit of the $\chi$SB
solutions of the compact model. Clearly this issue deserves to be
investigated further.

\section{\label{summary}Summary}

The study of $\chi$SB in QCD is made complicated by the fact that the
scale at which chiral symmetry is broken is of the order of the
confinement scale. If QCD could be deformed to enable ``tuning'' of
the $\chi$SB scale to be much smaller than the confinement scale, then
one would have separated the complications of the dynamics of
confinement from a study of $\chi$SB, which could then be handled by
perturbative methods. The intersecting brane configuration of Sakai
and Sugimoto, which gives rise to a QCD-like theory at low energies,
admits just such a possibility; it has an additional parameter, the
flavour brane-anti-brane separation, which can be tuned. In the strong
coupling limit, by tuning this parameter one can indeed raise the
chiral symmetry restoration temperature above the deconfinement
temperature \cite{ASY,PS0}. However, as we have seen in the present
work, for a consistent description of $\chi$SB in the weakly coupled
SS model, it is essential to incorporate the physics of
confinement. The interaction between flavour branes and anti-branes of
the SS model is captured by a non-local NJL model. For any finite
radius $R$ of the circle which the colour $D4$-branes wrap, there is
confinement and a mass gap in the low energy theory. The NJL model
reflects this dynamically generated mass scale, $\Lambda$, in the
length scale over which the non-local four-fermi interaction
extends. This fact turns out to be crucial in getting consistent
$\chi$SB solutions. In the large $N_c$ limit, the question of $\chi$SB
amounts to finding appropriate solutions to the non-linear gap
equation. For solutions with $\chi$SB length scale $l$ much larger
than the confinement scale $\Lambda^{-1}$, it is reasonable to replace
the non-local NJL model by the local NJL model. Hence these solutions
must reveal the existence of a critical coupling, which is known to
determine $\chi$SB in the local NJL model. In this paper we have
numerically solved the non-linear gap equation and verified the
existence of a critical coupling below which chiral symmetry is
unbroken. Roughly speaking, only solutions with $\chi$SB scale greater
than the brane-anti-brane separation $L$ exist. The $\chi$SB scale $l$
increases as the 't Hooft coupling is decreased, until a critical
coupling is reached for $l
\sim \Lambda^{-1}$. Solutions with $l >
\Lambda^{-1}$do not lead to any further decrease in the coupling.

Our analysis is valid for any finite value of the radius $R$, which
may be large. We have briefly addressed the question of what happens
when $R \rightarrow \infty$. Two different ways of taking this limit,
each one obtained from a well-motivated one-parameter scaling of the
parameters of the SS model, were discussed. We found from our
numerical data that neither of them leads to a sensible limit. The
tentative conclusion is that simple ways of implementing this limit do
not lead to a consistent picture of $\chi$SB in the non-compact
version of the non-local NJL model. This seems to reinforce the
critical role that the confinement scale plays in the compact model; the
infrared cut-off provided by it enables the existence of consistent
solutions to the gap equation. However, more work needs to be done to
clarify this issue further.

Finally, most of the calculations reported in this paper were done
numerically because the gap equation is non-linear and we could not
solve it analytically. It would, however, be useful to have some
analytic handle on the calculations, especially in the parameter
region near the critical coupling. This could be important for a
better understanding of the non-compact limit. A possible hint in this
respect is the fact that excellent numerical solutions were obtained
using the ansatz \eq{ansatz}, with the constant $c$ turning out to be
almost exactly equal to $l/L$ in all cases.

\bigskip

\newcommand{\sbibitem}[1]{\bibitem{#1}}

\end{document}